\documentclass[aps,prl,twocolumn,superscriptaddress,groupedaddress]{revtex4}
\usepackage{graphicx}
\usepackage{dcolumn}
\usepackage{bm}       
\usepackage{amssymb} 
\usepackage{amsmath}
\usepackage{amsfonts}
\usepackage{amsthm}
\usepackage{xfrac}
\newcommand{\be}{\begin{equation}}
\newcommand{\ee}{\end{equation}}
\newcommand{\ba}{\begin{align}}
\newcommand{\eal}{\end{align}}
\newcommand{\dg}{ ^{\dagger}}
\newcommand{\hml}{\mathcal{H}}

\begin{document}

\title{Quantum nonlinear dynamics of optomechanical systems in the strong coupling regime}
\author{J. D. P. Machado}
\author{Ya. M. Blanter}
\affiliation{Kavli Institute of Nanoscience, Delft University of Technology, Lorentzweg 1, 2628 CJ Delft, The Netherlands}

\date{\today}

\begin{abstract}
With an increasing coupling between light and mechanics, nonlinearities begin to play an important role in optomechanics. We solve the quantum dynamics of an optomechanical system in the multi-photon strong coupling regime retaining nonlinear terms. This is achieved by performing a Schrieffer-Wolff transformation on the Hamiltonian including driving terms. The approach is valid away from the red- and blue-sideband drive. We show that the mechanical resonator displays self-sustained oscillations in regimes where the linearized model predicts instabilities, and that the amplitude of these oscillations is limited by the nonlinear terms. Related oscillations of the photon number are present due to frequency mixing of the shifted mechanical and cavity frequencies. This leads to additional peaks in the cavity's power spectral density. Furthermore, we show that it is possible to create phonon states with sub-Poissonian statistics when the system is red-detuned. This result is valid even with strong driving and with initial coherent states.
\end{abstract}

\maketitle

By coupling light and mechanics, optomechanical systems enable control of light by mechanical motion and vice versa. This coupling of light to modes of a mechanical resonator is often achieved via radiation pressure. Its nature is intrinsically nonlinear but its strength is typically smaller than all other physical parameters \cite{bilbia}, undermining the significance of nonlinear effects. The majority of the observed physical phenomena can be thus understood using a linear description.

The achievement of cooling a mechanical mode to its ground state \cite{jackiechan, toifel} paves the way to state preparation and it increased the interest in quantum effects in mechanical motion, specifically creation of non-classical mechanical states. Such states can only be created in a nonlinear system, shifting the focus of attention to the single-photon strong coupling (SPSC) regime, where quantum dynamics and properties have been analyzed theoretically. The exact solution for the isolated system was discovered \cite{exact,exact_copy}. Dissipation, noise and the coherent drive have also been investigated by including them in the equations of motion and treating them as perturbations to the exact solution \cite{nunesko,photon blockade}. For this weak driving regime, it was found that mechanical states with sub-Poissonian statistics can be created in the SPSC regime. This conclusion was corroborated by numerical simulations \cite{elsevier,kubala,RAlame}. Other works reported similar quantum states \cite{coreia,quantumbs1, quantumbs2} over some parameter regions in the SPSC regime, and signalled a connection to the self-sustained oscillations present in the nonlinear regime. Experimentally, the SPSC regime remains a challenge, though considerable progress has been made.

Higher single-photon coupling strengths were already obtained in a variety of setups where the single-photon coupling surpasses the mechanical frequency $\Omega$ \cite{ewold} (or is close to it \cite{giga-coupling}). However, in these setups, the cavity linewidth exceeds the single-photon coupling.
Alternatively, the interaction can be enhanced by coherently driving the system. The drive enhances the single-photon coupling by a factor of $\sqrt{N_{photons}}$, which can go as high as $10^5$\cite{simoni} and may place this effective coupling higher than the cavity linewidth and close to the magnitude of the mechanical frequency \cite{simoni,simon, funny, toifel}, making this  multi-photon strong coupling (MPSC) regime more experimentally relevant. At the MPSC regime, fluctuations are usually negligible and linearizing the interaction is sufficient to describe the behaviour of the system. The resulting linear system can be solved, yielding instabilities for certain detunings. The instabilities can lead to the emergence of self-sustained oscillations. Such oscillations were already reported experimentally \cite{caos1} and for intense drive powers, chaotic motion takes place \cite{caos2}. To describe these phenomena, nonlinearities are essential, and in the absence of intrinsic nonlinearities, the nonlinear character of the interaction must be taken into account.
 The theoretical description of this regime with strong driving is mostly classical and so far, the effects of the quantum nonlinear interaction for the strong driving case could only be addressed recently using the Keldysh formalism approach \cite{marcantonio1,marcantonio2}.

In this Letter, we present a new approach to address the quantum dynamics of an optomechanical system that accounts for the nonlinearity of the interaction while considering strong driving. This approach is based on a Schrieffer-Wolff transformation \cite{olhaaonda} of the Hamiltonian and allows for an analytical solution. It also enables us to evaluate quantum properties of the system and to derive its dynamics in a consistent way without any a priori Ansatz. The results reveal the presence of self-sustained oscillations in the motion of the resonator and oscillations in the photon number arising from frequency mixing processes. These latter oscillations lead to the appereance of new peaks in the cavity's spectral density. In addition, we show that it is possible to create phonon states with sub-Poissonian statistics when the system is red-detuned. This result holds at the strong driving regime and even with initial coherent states. The approach is valid over a broad range of frequencies away from the red- and blue-detuned sidebands and for coupling strengths up to the mechanical frequency.

% COMECA AQUI

An optomechanical system can be modeled as two harmonic oscillators (the mechanical resonator and the cavity field) that couple through radiation pressure. This coupling is proportional to the field intensity and approximately linear in displacement.
Inclusion of the driving term leads to the Hamiltonian \cite{bilbia}:
\be
\hml=-\Delta a\dg a +\Omega b\dg b -g_0 a\dg a (b\dg +b)+\mathcal{E}(a\dg +a),
\ee
where $\Delta$ is the detuning of the driving laser from the cavity frequency $\omega_c$, $\Omega$ is the mechanical frequency, $g_0$ is the optomechanical coupling, $\mathcal{E}$ is the driving strength, and $a$ ($b$) represents the cavity's photon (phonon) annihilation operator in the frame rotating with the driving frequency $\omega_L$. In order to sweep away the direct influence of the driving term, a shift on the photon operator $a=A+\alpha$ is made,  with $\alpha$ chosen such that no driving terms appear on the equations of motion. For a weakly interacting system, $\alpha\approx \mathcal{E}/\big(\Delta+i\sfrac{k}{2}\big)$, where $k$ is the cavity linewidth. However, for the strong interacting case $\alpha$ must be obtained via a cubic equation (see Supplementary Information (SI)). With this shift, the operator $A$ obeys the same commutation relations as $a$ and represents the field displacements around the coherent component $\alpha$ produced by the driving laser. In the discussions ahead, the cavity states refer to the displaced states upon which $A$ acts. The Hamiltonian takes the form $\hml=\hml_0-\eta V$ with $\eta=g_0/\Omega$ and
\be
V=\Omega \Big[A\dg A(b\dg+b)+\alpha^*A(b+b\dg)+\alpha A\dg(b+b\dg)+|\alpha|^2(b+b\dg)\Big]\, .
\ee
For most experiments, the parameter $\eta$ is typically of the order of $10^{-3}$ or even less \cite{bilbia}. Thus, $V$ can be seen as a perturbation, and one can perform a unitary transformation to obtain an effective Hamiltonian for which the interaction is even weaker. This effective Hamiltonian is given by $U\hml U\dg$ with $U=e^{-\eta^2 T}e^{-\eta S}$, where $\{S,T\}$ are anti-Hermitian operators. After the transformation, the Hamiltonian takes the form
\be
\hml_{eff}=\hml_0-\eta^2\Big(\frac{1}{2}[S,V]+[T,\hml_0]\Big)+o(\eta^3),
\label{eq:htheo}
\ee
by imposing that $[S,\hml_0]=V$. This condition defines a Schrieffer-Wolff transformation, which removes the $1^{st}$ order dependence on $\eta$ and makes the effective interaction weaker. If one had included a bath in the Hamiltonian, an extra term of $\hml_{bath}$ would appear together with $\hml_0$. By doing so, all the noise and correlations with the bath could be obtained at any temperature, but at the cost of including an enormous amount of combinations of bath operators in $S$ and $T$ and compromising the solvability. Instead of including baths at the Hamiltonian level, damping terms for the light and mechanics are added to the equations of motion in order to account for dissipation.
There is a variety of possible choices for $\{S,T\}$ that lead to a Hamiltonian with the form of Eq.(\ref{eq:htheo}). However, most choices generate unwanted nonlinear terms. To avoid them, we choose $S$ and $T$ in the form
\ba
S=&d_1b +d_{11}Ab +d_{13}A\dg b+ d_{15}A\dg Ab-h.c. \nonumber \\
T=&d_{19}AA+d_{21} A\dg AA+d_{23}Abb+d_{25} A\dg bb  \nonumber \\
&+d_{27}Ab\dg b+d_{29}bb-h.c. \, ,
\end{align}
where h.c. stands for the Hermitian conjugate. The coefficients $\{d_{j}\}$ are shown in SI.
The explicit form of the effective Hamiltonian is
\be
\hml_{eff}=-\bar{\delta}A\dg A+\eta^2\Omega (A\dg A)^2+\tilde{\Omega}b\dg b \, ,
\label{eq:hamilniau}
\ee
and $\bar{\delta}$ and $\tilde{\Omega}$ are
\be
\bar{\delta}=\Delta+2\eta^2\Omega|\alpha|^2\frac{2\Omega^2-\Delta^2}{\Delta^2-\Omega^2}+4\eta^2Re\{(\mathcal{E}-\Delta \alpha)d_{21}\},
\ee
\be
\tilde{\Omega}=\Omega-2\eta^2\Omega|\alpha|^2\frac{\Delta\Omega}{\Delta^2-\Omega^2}-2\eta^2Re\{(\mathcal{E}-\Delta \alpha)d_{27}\}.
\label{eq:omiga}
\ee
The only approximation considered was to disregard high order terms (h.o.t.) in $\eta$. Although $\eta$ is extremely small, the h.o.t contain powers of $\alpha$, so they might be non-negligible for certain parameters. The evaluation of the h.o.t. establishes the critical value $g_{crit}$ at which they begin to play a role. This $g_{crit}$ is independent of the cavity linewidth $k$ and it is always smaller than the mechanical frequency. Additionally, $g_{crit}\to 0$ at $\Delta=\{0,\pm \Omega\}$, which excludes these detunings from the validity range, and the vicinity of these points can only be accessed for very small $g:=g_0|\alpha|$. 
A deeper analysis of the dependence of $g_{crit}$ on the physical parameters is shown in SI.

With Eq.(\ref{eq:hamilniau}), obtaining the solution to the equations of motion is straightforward. The time-evolution for the operators in the basis that diagonalizes the Hamiltonian, $\{A_N,b_N\}$, is given by
\be
 A_N(t)=e^{-i\phi(t)}A_N(0)\quad\textbf{,}\quad
b_N(t)=e^{-i\tilde{\Omega}t-\frac{\Gamma}{2}t}b_N(0) \, ,
\label{eq:sols}
\ee
where $\Gamma$ is the mechanical damping rate and
\be
\phi(t)=\Big(-\bar{\delta}-i\frac{k}{2}\Big)t -2\frac{\Omega}{k}\eta^2 (e^{-kt}-1)(A_N\dg A_N) (0) .
\ee
These solutions immediately display known phenomena, such as the optical spring effect (visible in Eq.(\ref{eq:omiga})) and the avoided crossing at $\Delta=-\Omega$. The avoided crossing can be seen by evaluating $\tilde{\Omega}$ and $\bar{\delta}$ at $\Delta=-\Omega+\zeta$ and verifying that close to $\zeta=0$, the gap separating the frequencies is $\tilde{\Omega}+\bar{\delta}\approx 2\sfrac{g^2}{\zeta}+o(\zeta)$.

To evaluate physical observables, we express the operators in the original basis that represents the physical entities: $\{a_o,b_o\}$. They are related to $\{A_N,b_N\}$ by
\be
a_{o}=U\dg A_NU+\alpha \quad \textbf{;}\quad b_{o}=U\dg b_NU.
\label{eq:carlosrocha}
\ee
A closed form for the representation of the operators in the original basis is impossible to obtain. Nevertheless, one can estimate the expressions at Eq.(\ref{eq:carlosrocha}) by expanding $U$ in powers of $\eta$ and arrive at simple expressions relating the new basis and the original one.
% Thus, an operator $O$ transform as
%\ba
%O_o  \approx O_N+\eta [S,O_N]+\eta^2\big([T,O_N]+\frac{1}{2}[S,[S,O_N]]\big) .
%\end{align}
Next, we focus on the strong driving regime, defined by $|\alpha|^2\gg \eta^{-1}\sqrt{N_b}$, with $N_b$ the mean phonon number. This regime makes the effective interaction important. It also simplifies the expressions, since some terms for the time-evolution of the fields become negligible.\\
\textbf{Cavity field.}
The time-evolution for the cavity field $A_o$ in the strong driving regime is given by
\ba
A_o(t)&\approx \Upsilon_0(t)+\Upsilon_1(t)A_o(0)+\Upsilon_2(t)A_o\dg (0)\nonumber \\
&+\Upsilon_5(t) A_o\dg (0)A_o (0)+\Upsilon_6(t) A_o(0)A_o(0)\nonumber\\
&+\Upsilon_{10}(t) A_o\dg (0)A_o(0)A_o(0) .
\label{eq:A}
\end{align}
The $\{\Upsilon_i\}$ terms are composed of combinations of oscillating terms, whose frequencies are the shifted frequencies $\bar{\delta} , \tilde{\Omega}$ and combinations of these frequencies (such as $\bar{\delta}+\tilde{\Omega}, \bar{\delta}-\tilde{\Omega}, 2\bar{\delta}$), and their expressions are on SI.
With Eq.(\ref{eq:A}), the statistics for the cavity field can be evaluated. Due to the strong coherent driving, the cavity field acquires a coherent component and commutation relations between the operators may have a negligible effect. The state for which the commutations are most relevant is the $|0\rangle$ state. For it, the $2^{nd}$ order correlation function takes the value
\be
g^{(2)}\approx 1+4\bigg(\frac{g}{\Omega}\bigg)^4\frac{1}{|\alpha|^2}\approx 1
\ee
in the strong driving limit. Evaluating any other correlation function would render a similar result, i.e., the corresponding value for a coherent state.
As the photon state can be obtained through evaluation of all correlation functions, the cavity field must be in a coherent state at all times \footnote{for more details see (SI)}. It must be also noted that any difference between the average values of operators with a different order will be of the same order of magnitude as the previously disregarded terms. This classicality stems from the strong coherent driving, which imposes a large coherent state for the cavity field.

\begin{figure}[h]
\includegraphics[scale=0.65]{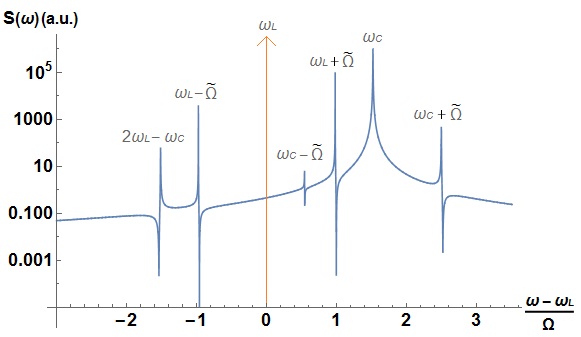}
\caption{Power spectral density of the cavity field for $\Delta=-1.5\Omega$, $g=0.1\Omega$ and $k=0.001\Omega$. The nonlinear interaction generates sidebands around the cavity frequency. A peak at $2\omega_L-\omega_c-\tilde{\Omega}$ exists but it is not visible for these parameters.\label{fig:picos}}
\end{figure}

From the time-dependence of $\{\Upsilon_i\}$, it can be seen that the cavity field exhibits frequencies which are combinations of the cavity, the laser and the mechanical frequencies. The focus of attention is commonly towards the cavity frequency and the sidebands around the laser frequency \cite{bilbia}. In the MPSC regime, additional frequencies arise, such as sidebands around the cavity frequency, and they are signatures of the nonlinear interaction.
By evaluating the power spectral density for the cavity field, one can see the appearance of these new frequencies. In order for them to be visible, one must be on the resolved sideband regime ($\Omega\gg k$) and choose a proper detuning such that the peaks do not overlap. These frequencies are often present but for some initial states, some of them vanish. Let us denote the initial states as $|\varphi ;\mu\rangle$, where $|\varphi\rangle$ is the cavity state and $|\mu\rangle$ is the mechanical state. The spectral density for the cavity and for the initial state $|-1.1\alpha ; 0\rangle$ is exhibited on Fig.\ref{fig:picos}. 
Note that the multiple frequencies were reported in \cite{ewold,caos2}, but the underlying mechanisms are different from the one discussed here.

\begin{figure}[h]
\includegraphics[scale=0.65]{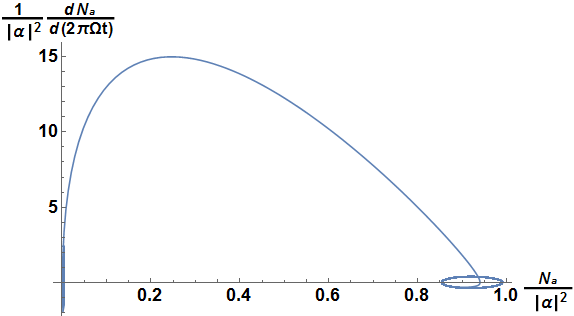}
\caption{Photon number time evolution for $\Delta=-1.8\Omega$, $g=0.3\Omega$ and $k=10\Omega$. The system will converge to an ellipse in $(N_a, d_tN_a)$ phase space, which represents an oscillation of the photon number $\langle a\dg a\rangle$.\label{fig:bola}}
\end{figure}

\begin{figure}[h]
\includegraphics[scale=0.65]{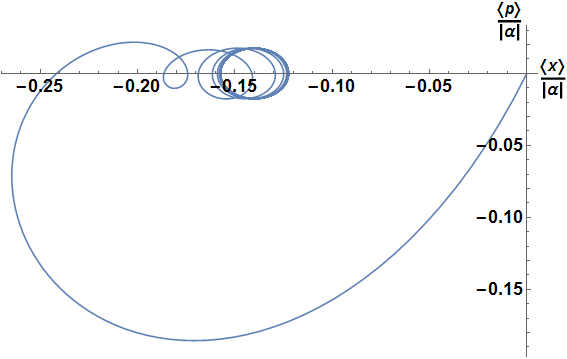}
\caption{Time-evolution of the resonator in the (x,p) phase space for $\Delta=0,6\Omega$, $k=\Omega$, $g=0,07\Omega$. For these parameters, the linearized model predicts an instability. The initial growth of the displacement converges to an ellipse in phase space, which represents self-sustained oscillations. \label{fig:a4}}
\end{figure}

As the coupling increases, so does the amplitude of the oscillations at these frequencies, and the oscillations become prominent. They are responsible for the appearance of oscillations in the photon number. The time-evolution of the photon number can be described better by representing the trajectory in the $(N_a, d_tN_a)$ phase space. By doing so, one can visualize the whole time-evolution and identify the limit cycles.
Evaluating the time-evolution of the photon number $\langle a\dg a\rangle(t)$ with the same initial state, one can see in Fig.\ref{fig:bola} that the motion of the system converges to an ellipse in the $(N_a, d_tN_a)$ phase space, representing the oscillation in the photon number.
\\
Photon number oscillations are known to be connected to self-sustained oscillations of the mechanical resonator \cite{caos2}, and we can address the time-evolution for the displacement of the resonator in the nonlinear regime without any a priori assumption.\\
\textbf{Mechanical resonator.}
The time-evolution of $b_o(t)$ is
\ba
b_o(t)\approx& \Xi_0(t)+\Xi_4(t)b_0(0)+\Xi_5(t)b_o\dg(0)+\Xi_1(t)A_o(0)\nonumber \\
&+\Xi_2(t)A_o\dg (0)+\Xi_3(t) A_o\dg (0)A_o (0) ,
\label{eq:b}
\end{align}
and the expressions for $\{\Xi_j\}$ are in SI.
%\ba
%\Xi_0(t)=& -\eta|\alpha|^2(1-e^{-\frac{\Gamma}{2}t-i\tilde{\Omega}t}) \, ;\\
%\Xi_1(t)=&\eta d_{13}^*(e^{-\frac{k}{2}t-i\phi (t)}-e^{-\frac{\Gamma}{2}t-i\tilde{\Omega}t}) \, ;\\
%\Xi_2(t)=&\eta d_{11}^*(e^{-\frac{k}{2}t+i\phi (t)}-e^{-\frac{\Gamma}{2}t-i\tilde{\Omega}t})   \, ;\\
%\Xi_3(t)=&\eta (e^{-\frac{\Gamma}{2}t-i\tilde{\Omega}t}-e^{-kt}) \, .
%\end{align}
It is known from the linearized model that for some parameters a negative effective damping occurs, leading to an instability. This instability can arise in the blue sideband, even in the weak coupling regime. From Eq.(\ref{eq:b}), the time-evolution for the displacement of the resonator $\langle x(t)\rangle =\langle b(t)+b\dg (t)\rangle$ is easily obtained, and with it, the complete dynamics past the initial instability can be evaluated. Considering the initial state $|-\alpha ;0\rangle$, it is seen from Fig.\ref{fig:a4}, that after the initial growth of displacement (lasting $k^{-1}$), the resonator starts to develop long-lasting oscillations . This motion of the mechanical resonator is a characteristic of self-sustained oscillations, which can be identified by the elliptic shape drawn by the motion in phase space. The origin of these oscillations can be found in the expression for $\langle x(t)\rangle $ and it arises from destructive interference between two oscillations with different decay rates.

\begin{figure}[h]
\includegraphics[scale=0.6]{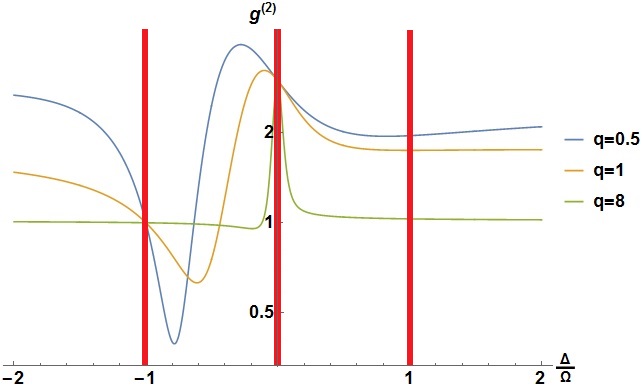}
\caption{$g^{(2)}(t^*_m)$ as a function of detuning for the initial state $|q;0\rangle$, with the coherent photon state $|q\rangle =\{|0.5\rangle,|1\rangle,|8\rangle\}$. The mechanical state exhibits sub(super)-Poissonian statistics when red(blue)-detuned for initial states with $|q|\leq 1$. As the coherent states become larger, the effect vanishes. It is assumed that the transient time $k^{-1}$ has past. The red shaded regions denote where the approach does not hold.\label{fig:g2}}
\end{figure}

 With Eq.(\ref{eq:b}), the phonon statistics can be computed. In the strong driving regime, the mechanical resonator acquires a massive coherent component, and by a similar reasoning used for the cavity field, the phonons are in a coherent state most of the times. However, the large coherent component $\Xi_0$ vanishes periodically every $t_m^*=\frac{2\pi m}{\tilde{\Omega}}$, in the limit $\Gamma\to 0$.
This allows a short time window ($\Delta \tau\propto |\alpha|^{-1}$), when the resonator is able to depart from a coherent state. If the resonator starts in the ground state, the coupling does not play a role in the state developed at $t_m^*$, and after the transient time $k^{-1}$, the state is determined solely by the detuning and the initial cavity state.  For an initial state $|q ;0\rangle$, where $|q\rangle$ is a coherent state, sub-Poissonian statistics are visible when red-detuned  (see Fig.\ref{fig:g2}) and when $|q|\leq 1$. As the amplitude of the initial coherent state increases, this effect begins to disappear.
Thus, it is possible for the phonon state to have sub-Poissonian statistics at the particular times $t^*_m$ if the initial state is a small coherent photon state and the mechanical resonator is in the ground state. This is not restricted to these initial states. Yet, it is remarkable that there is a possibility to create states with sub-Poissonian statistics starting with an initial coherent state and driving the system with a strong coherent drive. More details about the necessary conditions can be found in SI.

Summarizing, we have shown that resorting to a Schrieffer-Wolff transformation, it is possible to obtain analytical solutions for the quantum dynamics of a strongly driven optomechanical system when the nonlinear interaction plays an important role. With this approach, the time-evolution of the cavity and mechanical fields were obtained using an operator description, thus enabling us to evaluate their quantum properties. 

The solutions show that the mechanical resonator develops self-sustained oscillations for parameter regions where the linearized model predicts an instability. Thus, the nonlinear interaction limits the growth produced by a negative effective damping. Similar oscillations are also present in the photon number. Their origin is the existence of distinct frequencies in the cavity field, as revealed by the power spectral density. 

We have also shown that it is possible to create phonon states with sub-Poissonian statistics if the system is red-detuned. This result holds even in the strong driving regime and with initial (small) coherent states.

\end{document}